\documentclass[aps,prd,preprint,showpacs,preprintnumbers,nofootinbib,superscriptaddress]{revtex4}
\usepackage{ae}
\usepackage{bm} 
\usepackage{color}
\usepackage{amsmath}
\usepackage{amssymb}
\usepackage{amsfonts}
\usepackage{graphicx}
\usepackage{slashed}
\usepackage{wasysym}
\usepackage{setspace}

\newcommand{\Eqref}[1]{Eq.~\eqref{#1}}

\allowdisplaybreaks

\begin{document}

\setlength{\unitlength}{1mm}
\title{Photon propagation in slowly varying inhomogeneous electromagnetic fields}
\author{Felix Karbstein}\email{felix.karbstein@uni-jena.de}
\affiliation{Helmholtz-Institut Jena, Fr\"obelstieg 3, 07743 Jena, Germany}
\affiliation{Theoretisch-Physikalisches Institut, Abbe Center of Photonics, \\ Friedrich-Schiller-Universit\"at Jena, Max-Wien-Platz 1, 07743 Jena, Germany}
\author{Rashid Shaisultanov}\email{shaisultanov@gmail.com}
\affiliation{Nazarbayev University, NURIS block 9, Laboratory of Nano Synergy, 53 Kabanbay Batyr Ave., Astana, 010000, Republic of Kazakhstan}

\date{\today}

\begin{abstract}
 Starting from the Heisenberg-Euler effective Lagrangian, we determine the photon current and photon polarization tensor in inhomogeneous, slowly varying electromagnetic fields. 
 To this end, we consider background field configurations varying in both space and time, paying special attention to the tensor structure.
 As a main result, we obtain compact analytical expressions for the photon polarization tensor in realistic Gaussian laser pulses, as generated in the focal spots of high-intensity lasers.
 These expressions are of utmost importance for the investigation of quantum vacuum nonlinearities in realistic high-intensity laser experiments.
\end{abstract}


\maketitle

\section{Introduction} \label{sec:intro}

As has been recognized already in the early days of quantum electrodynamics (QED), the fluctuations of virtual charged particles in the vacuum  give rise to nonlinear, effective couplings between electromagnetic fields \cite{Euler:1935zz,Heisenberg:1935qt,Weisskopf}.
However, so far the pure electromagnetic nonlinearity of the quantum vacuum though subject to high-energy experiments \cite{Akhmadaliev:1998zz} has not been directly verified on macroscopic scales.
The advent and planning of high-intensity laser facilities of the petawatt class has triggered a huge interest in proposals to probe quantum vacuum nonlinearities in realistic all-optical experimental set-ups;
for recent reviews, see \cite{Dittrich:2000zu,Marklund:2008gj,Dunne:2008kc,Heinzl:2008an,DiPiazza:2011tq}.
Prominent examples are proposals intended to verify vacuum birefringence \cite{Toll:1952,Baier,BialynickaBirula:1970vy,Adler:1971wn} with the aid of high-intensity lasers \cite{Heinzl:2006xc} (cf. also \cite{Dinu:2013gaa}),
so far searched for in experiments using macroscopic magnetic fields \cite{Cantatore:2008zz,Berceau:2011zz}.
Complementary suggestions promote the use of time-varying fields and high-precision interferometry \cite{Zavattini:2008cr,Dobrich:2009kd,Grote:2014hja}.
Other commonly studied nonlinear vacuum effects encompass direct light-by-light scattering \cite{Euler:1935zz,Karplus:1950zz}, photon splitting \cite{Adler:1971wn}, and spontaneous vacuum decay in terms
of Schwinger pair-production in electric fields \cite{Sauter:1931zz,Heisenberg:1935qt,Schwinger:1951nm}.
Besides, recent theoretical studies have focused on optical signatures of quantum vacuum nonlinearities based on interference effects \cite{King:2013am,Tommasini:2010fb,Hatsagortsyan:2011}, photon-photon scattering in the form of laser-pulse collisions \cite{King:2012aw}, quantum reflection \cite{Gies:2013yxa}, photon merging \cite{Gies:2014jia}, and harmonic generation from laser-driven vacuum \cite{DiPiazza:2005jc,Fedotov:2006ii}.

A central object in the study of such effects is the photon polarization tensor, which is known analytically in several limits, namely for homogeneous electric and/or magnetic fields \cite{BatShab,narozhnyi:1968,ritus:1972,Tsai:1974fa,Tsai:1974ap,Baier:1974hn,Urrutia:1977xb,Dittrich:2000wz,Schubert:2000yt}
(cf. also \cite{Dittrich:2000zu,Karbstein:2013ufa} and references therein), and generic plane wave backgrounds
\cite{Baier:1975ff,Becker:1974en}; see \cite{Meuren:2013oya} for a more recent derivation and an alternative representation, and
\cite{Gies:2014jia} for a novel systematic expansion especially suited for all-optical experimental scenarios.
Numerical results for inhomogeneous magnetic backgrounds are available from worldline Monte Carlo simulations \cite{Gies:2011he},
and first analytical insights into the photon polarization tensor in various inhomogeneous background field configurations that may locally be approximated by a constant were obtained in \cite{Gies:2013yxa}.
However, the latter approach is manifestly limited to certain photon polarization modes as the incorporation of the inhomogeneous field profile into the constant field polarization tensor generically induced violations of the Ward identity.

In this paper we study probe photon propagation in generic inhomogeneous electromagnetic fields varying in both space and time.
In contrast to previous studies which implicitly assumed constant background fields when extracting the photon polarization tensor from the Heisenberg-Euler Lagrangian (cf., e.g., \cite{BialynickaBirula:1970vy,Heinzl:2006xc}), we treat the background field as manifestly inhomogeneous from the outset.
Our main interest is on analytical insights beyond the explicitly known field configurations, i.e., beyond constant and homogeneous and plane wave background fields.

\section{Low-energy effective theory of photon propagation} \label{sec:mainpart}

At one-loop order, the effective Lagrangian in constant
external electromagnetic fields (``Heisenberg-Euler effective Lagrangian'') \cite{Heisenberg:1935qt},
describing the effective nonlinear interactions between electromagnetic fields mediated by electron-positron fluctuations in the vacuum,
can be represented concisely in terms of the following propertime integral \cite{Schwinger:1951nm} (cf. also \cite{Dittrich:2000zu,Jentschura:2001qr}),
\begin{equation}
 {\cal L}
 =\frac{\alpha}{2\pi}\int_{0}^{\infty}\frac{{\rm d}s}{s}\,{\rm e}^{-i\frac{m^2}{e}s}\left[ab\coth(as)\cot(bs)-\frac{a^2-b^2}{3}
 -\frac{1}{s^2}\right], \label{eq:effL}
\end{equation}
with the prescription $m^2\to m^2-i0^+$, and the propertime integration contour assumed to lie slightly below the real positive $s$ axis.
Here, $m$ is the electron mass, $e$ is the elementary charge, $\alpha=e^2/(4\pi)$ is the fine-structure constant, 
and $a=(\sqrt{{\cal F}^2+{\cal G}^2}-{\cal F})^{1/2}$ and $b=(\sqrt{{\cal F}^2+{\cal G}^2}+{\cal F})^{1/2}$ are the secular invariants made up of the gauge and Lorentz invariants of the electromagnetic field: ${\cal F}=\frac{1}{4}F_{\mu\nu}F^{\mu\nu}=\frac{1}{2}(\vec{B}^2-\vec{E}^2)$ and ${\cal G}=\frac{1}{4}F_{\mu\nu}{}^*F^{\mu\nu}=-\vec{E}\cdot\vec{B}$,
with $^*F^{\mu\nu}=\frac{1}{2}\epsilon^{\mu\nu\alpha\beta}F_{\alpha\beta}$ denoting the dual field strength tensor; $\epsilon^{\mu\nu\alpha\beta}$ is the totally antisymmetric tensor, fulfilling $\epsilon^{0123}=1$.
Our metric convention is $g_{\mu \nu}=\mathrm{diag}(-1,+1,+1,+1)$, and we use units where $c=\hbar=1$.
To keep notations compact we moreover employ the shorthand notations
$\int_x\equiv\int{\rm d}^4x$ and $\int_k\equiv\int\frac{{\rm d}^4k}{(2\pi)^4}$
for the integrations over position and momentum space, respectively.

The effective Lagrangian~\eqref{eq:effL} is a gauge and Lorentz invariant quantity.
Clearly, for inhomogeneous background fields additional gauge and Lorentz invariant building blocks become available.
For slowly varying fields the deviations from the constant field limit can be accounted for with derivative terms $\sim\partial_\alpha F^{\mu\nu}$.
If the typical frequency/momentum scale of variation of the inhomogeneous background field is $\upsilon$, derivatives effectively translate into multiplications with $\upsilon$ to be rendered dimensionless by the electron mass $m$.
Thus, \Eqref{eq:effL} is also applicable for slowly varying inhomogeneous fields fulfilling $\frac{\upsilon}{m}\ll1$, or -- in other words -- for inhomogeneities whose typical spatial (temporal) scales of variation are much larger than the Compton wavelength (time) $\sim\frac{1}{m}$ of the virtual charged particles.
The electron Compton wavelength is $\lambda_c=3.86\cdot10^{-13}{\rm m}$ and the Compton time is $\tau_c=1.29\cdot10^{-21}{\rm s}$.
In turn, many electromagnetic fields available in the laboratory, e.g., the electromagnetic field pulses generated by optical high-intensity lasers, featuring wavelengths of ${\cal O}(\mu{\rm m})$ and pulse durations of ${\cal O}({\rm fs})$, are compatible with this requirement.
The effective Lagrangian is a scalar quantity, and scalar quantities made up of combinations of $F^{\mu\nu}$, ${}^*F^{\mu\nu}$ and derivatives thereof involve an even number of derivatives.
Hence, when employing the constant field result~\eqref{eq:effL} for slowly varying inhomogeneous fields, the deviations from the corresponding (unknown) exact result are of ${\cal O}\bigl((\tfrac{\upsilon}{m})^2\bigr)$.

\pagebreak

Within the above restrictions, \Eqref{eq:effL} can serve as a starting point to study the effective interaction between dynamical photons and inhomogeneous background electromagnetic fields.
For this purpose it is convenient to decompose the electromagnetic field strength tensor $F^{\mu\nu}$ introduced above as $F^{\mu\nu}\to F^{\mu\nu}(x)+f^{\mu\nu}(x)$ into the field strength tensor of the background field $F^{\mu\nu}(x)=\partial^\mu A^\nu(x) - \partial^\nu A^\mu(x)$ and the photon field strength tensor $f^{\mu\nu}(x)=\partial^\mu a^\nu(x) - \partial^\nu a^\mu(x)$ \cite{BialynickaBirula:1970vy}.
The effective action can then be compactly written as $S=S_{\rm MW}+S_{\rm int}$,
where $S_{\rm MW}=-\frac{1}{4}\int_x\mathfrak{F}^{\mu\nu}(x)\mathfrak{F}_{\mu\nu}(x)$ is the Maxwell action of classical electrodynamics, with $\mathfrak{F}^{\mu\nu}(x)\equiv F^{\mu\nu}(x)+f^{\mu\nu}(x)$ denoting the field strength of both the background and the dynamical electromagnetic fields.
The additional interaction term $S_{\rm int}=\sum_{n=1}^\infty S_{\rm int}^{(n)}$ encodes quantum corrections and vanishes in the limit $\hbar\to0$.
It can be expanded in terms of interactions involving $n$ photons, i.e., $S_{\rm int}^{(n)}\sim f^n$, with $f\equiv f^{\mu\nu}$.
In particular, to quadratic order in $f$ it is given by $S_{\rm int}=S_{\rm int}^{(1)}+S_{\rm int}^{(2)}+{\cal O}(f^3)$, with
\begin{equation}
 S_{\rm int}^{(1)} = \int_x f^{\mu\nu}(x)\frac{\partial{\cal L}}{\partial F^{\mu\nu}}(x) \quad\text{and}\quad
 S_{\rm int}^{(2)} = \frac{1}{2}\int_x f^{\alpha\beta}(x)\frac{\partial^2{\cal L}}{\partial F^{\alpha\beta}\partial F^{\mu\nu}}(x)f^{\mu\nu}(x) \,.
\end{equation}
The neglected higher-order terms of ${\cal O}(f^3)$ correspond to interactions involving three or more photons.
For a somewhat similar approach in a completely different context, cf. \cite{Shaisultanov:1997bc,Karbstein:2007be}.

Due to the fact that the effective Lagrangian~\eqref{eq:effL} is local and thus depends only on a single position space coordinate $x$, all the effective interactions to be induced in the limit of slowly varying fields will be local,
or -- attributing some nonlocality to the spatial derivatives -- at least ``almost local''.

Equation~\eqref{eq:effL} is straightforwardly differentiated with respect to $F^{\mu\nu}$, and the tensor structure of the resulting expression can be spanned by $F_{\mu\nu}$ and ${}^*F_{\mu\nu}$ (cf. also \cite{Karbstein:2014fva}).
We find it convenient to write it as
\begin{equation}
 \frac{\partial{\cal L}}{\partial F^{\mu\nu}}=\frac{1}{2}\biggl( \frac{\partial {\cal L}}{\partial{\cal F}} F_{\mu\nu} +  \frac{\partial{\cal L}}{\partial{\cal G}}\,{}^*F_{\mu\nu}\biggr). \label{eq:dL/dF}
\end{equation}
To keep notations simple, we have omitted any explicit reference to the $x$ dependence of the electromagnetic fields.
The tensor structure of the second derivative is slightly more complicated. It is spanned by six independent tensor structures and reads
\begin{multline}
 \frac{\partial^2{\cal L}}{\partial F^{\alpha\beta}\partial F^{\mu\nu}}
 =\frac{1}{4}\biggl[\bigl(g_{\alpha\mu}g_{\beta\nu}-g_{\alpha\nu}g_{\beta\mu}\bigr)\frac{\partial\cal L}{\partial{\cal F}}  +  \epsilon_{\mu\nu\alpha\beta}\, \frac{\partial\cal L}{\partial{\cal G}} 
 + F_{\mu\nu}F_{\alpha\beta}\frac{\partial^2 {\cal L}}{\partial{\cal F}^2} +  {}^*F_{\mu\nu}{}^*F_{\alpha\beta}\frac{\partial^2 {\cal L}}{\partial{\cal G}^2} \\
 + \bigl(F_{\mu\nu}{}^*F_{\alpha\beta} + {}^*F_{\mu\nu}F_{\alpha\beta}\bigr)\frac{\partial^2 {\cal L}}{\partial{\cal F}\partial{\cal G}}\biggr].
 \label{eq:ddL/dFsF}
\end{multline}
For the reasons given above for the effective Lagrangian, \Eqref{eq:ddL/dFsF} is valid up to corrections of ${\cal O}\bigl((\frac{\upsilon}{m})^2\bigr)$.

\pagebreak

Turning to the Fourier representation of the photon field, $a^\mu(x)=\int_k {\rm e}^{ikx} a^\mu(k)$, we write the probe photon field strength tensor as
$f^{\mu\nu}(x)= i\int_k {\rm e}^{ikx} \bigl[k^\mu g^{\nu\sigma} - k^\nu g^{\mu\sigma}\bigr]a_\sigma(k)$.
With the help of this expression,
and taking into account that ${\cal L}$ is localized, i.e., vanishes at large $x$, such that we can safely interchange the momentum and space integrations,
we obtain
\begin{gather}
 S_{\rm int}^{(1)} = \int_k j^\sigma(k)a_\sigma(k) \quad\text{and}\quad
 S_{\rm int}^{(2)} = -\frac{1}{2}\int_{k}\int_{k'} a_\rho(k)\Pi^{\rho\sigma}(k,k')a_\sigma(k')\,,
\end{gather}
where we have introduced the {\it photon current}
\begin{equation}
 j^\sigma(k) \equiv\frac{\delta S_{\rm int}}{\delta a_\sigma(k)}\biggl|_{a=0} = i\bigl(k^\mu g^{\nu\sigma} - k^\nu g^{\mu\sigma}\bigr)\int_x{\rm e}^{ikx}\frac{\partial{\cal L}}{\partial F^{\mu\nu}}(x) \label{eq:j0}
\end{equation}
and the {\it photon polarization tensor} in momentum space
\begin{multline}
 \Pi^{\rho\sigma}(k,k') \equiv -\frac{\delta^2 S_{\rm int}}{\delta a_\rho(k)a_\sigma(k')}\biggl|_{a=0} \\
 = \bigl(g^{\rho\beta}k^\alpha  - g^{\rho\alpha}k^\beta \bigr)
  \biggl[\int_x{\rm e}^{i(k+k')x}\frac{\partial^2{\cal L}}{\partial F^{\alpha\beta}\partial F^{\mu\nu}}(x)\biggr]
  \bigl(k'^\mu g^{\nu\sigma} - k'^\nu g^{\mu\sigma}\bigr) . \label{eq:Pi0}
\end{multline}
For completeness note that the corresponding polarization tensor in position space, obtained by a Fourier transform of \Eqref{eq:Pi0}, can be represented in an analogous fashion,
\begin{multline}
 \Pi^{\rho\sigma}(x,x') \equiv \frac{\delta^2 S_{\rm int}}{\delta a_\rho(x)a_\sigma(x')}\biggl|_{a=0} \\
 =\bigl(g^{\rho\beta}\partial_x^\alpha  - g^{\rho\alpha}\partial_x^\beta \bigr)\bigl(g^{\nu\sigma}\partial_{x'}^\mu - g^{\mu\sigma}\partial_{x'}^\nu\bigr)\biggl[\delta(x-x')\frac{\partial^2{\cal L}}{\partial F^{\alpha\beta}\partial F^{\mu\nu}}(x)\biggr].
\end{multline}
Tensors of higher rank, describing the effective interaction of $n$ photons can be derived along the same lines.

As the field strength tensor $f^{\mu\nu}$ of the probe photon field in momentum space is linear in $k^\sigma$, the photon polarization tensor in momentum space is at least quadratic in $k^\sigma$.
Taking into account that our approach is limited to slowly varying fields from the outset, we have ${\cal O}(k)={\cal O}(\upsilon)$.
Hence, the neglected contributions to the polarization tensor are $\sim\upsilon^2\,{\cal O}\bigl((\frac{\upsilon}{m})^2\bigr)$.
Employing the shorthand notation $(kF)^\mu=k_\nu F^{\nu\mu}$, $(k{}^*F)^\mu=k_\nu{}^*F^{\nu\mu}$, $(kk')=k_\mu k'^\mu$, etc.,
upon insertion of Eqs.~\eqref{eq:dL/dF} and \eqref{eq:ddL/dFsF} into Eqs.~\eqref{eq:j0} and \eqref{eq:Pi0} we obtain
\begin{equation}
 j^\sigma(k) = i\int_x{\rm e}^{ikx} \biggl[ (kF)^\sigma\,\frac{\partial{\cal L}}{\partial{\cal F}}
 + (k{}^*F)^\sigma\,\frac{\partial{\cal L}}{\partial{\cal G}}\biggr] \label{eq:j}
\end{equation}
and
\begin{multline}
 \Pi^{\rho\sigma}(k,k')
 =\int_x{\rm e}^{i(k+k')x} \biggl[
 \bigl((k k')g^{\rho\sigma} - k'^\rho k^\sigma \bigr)\frac{\partial\cal L}{\partial{\cal F}}
 + k'_\mu k_\alpha \epsilon^{\rho\sigma\mu\alpha}\, \frac{\partial\cal L}{\partial{\cal G}} \\
 + (kF)^\rho  (k'F)^\sigma\,\frac{\partial^2 {\cal L}}{\partial{\cal F}^2}
 + (k{}^*F)^\rho (k'{}^*F)^\sigma\,\frac{\partial^2 {\cal L}}{\partial{\cal G}^2} \\
 + \bigl[(k{}^*F)^\rho (k'F)^\sigma + (k F)^\rho (k'{}^*F)^\sigma\bigr]\,\frac{\partial^2 {\cal L}}{\partial{\cal F}\partial{\cal G}}
 \biggr]. \label{eq:Pi}
\end{multline}
Note that $k_\sigma j^\sigma(k)=0$ and $k_\rho\Pi^{\rho\sigma}(k,k')=\Pi^{\rho\sigma}(k,k')k'_\sigma=0$, i.e., the Ward identity is fulfilled, and correspondingly gauge invariance with respect to gauge transformations of the probe photon field is ensured.
In the limit of constant background fields, the integrands in Eqs.~\eqref{eq:Pi0} and \eqref{eq:Pi} do not depend on $x$ apart from the overall phase factor ${\rm e}^{i(k+k')x}$, such that
the integration over $x$ can be performed right away resulting in an overall momentum conserving Dirac delta function $\sim\delta(k+k')$.
This reproduces the known expression of the photon polarization tensor in the constant field and small momentum limit \cite{Shabad:2011hf}.

The five different derivatives of the effective Lagrangian~\eqref{eq:effL} with respect to ${\cal F}$ and ${\cal G}$ appearing in Eqs.~\eqref{eq:j}-\eqref{eq:Pi} can straightforwardly be taken,
and their explicit propertime integral representations read
\begin{align}
 \frac{\partial{\cal L}}{\partial{\cal F}}
 =&\frac{\alpha}{2\pi}\int_{0}^{\infty}\frac{{\rm d}s}{s}\,{\rm e}^{-i\frac{m^2}{e}s}\biggl\{\frac{ab}{a^2+b^2}\frac{as\cot(bs)}{\sinh^2(as)}+(a\leftrightarrow ib)+\frac{2}{3}\biggr\},
\nonumber\\
 \frac{\partial{\cal L}}{\partial{\cal G}}
 =&\frac{\alpha}{2\pi}\int_{0}^{\infty}\frac{{\rm d}s}{s}\,{\rm e}^{-i\frac{m^2}{e}s}\,
 {\cal G}\biggl\{\frac{\coth(as)\cot(bs)}{2ab} -\frac{1}{a^2+b^2}\frac{bs\cot(bs)}{\sinh^2(as)} + (a\leftrightarrow ib) \biggr\},
\nonumber\\
 \frac{\partial^2{\cal L}}{\partial{\cal F}^2}
 =&\frac{\alpha}{2\pi}\int_{0}^{\infty}\frac{{\rm d}s}{s}\,{\rm e}^{-i\frac{m^2}{e}s} \frac{ab}{(a^2+b^2)^2} 
 \biggl\{-\frac{abs^2}{\sinh^2(as)\sin^2(bs)} \nonumber\\
 &\hspace*{2cm}-\Bigl[1-2as\coth(as)-2\frac{a^2-b^2}{a^2+b^2}\Bigr]\frac{as\cot(bs)}{\sinh^2(as)}  + (a\leftrightarrow ib)
  \biggr\},
\nonumber\\
 \frac{\partial^2{\cal L}}{\partial{\cal G}^2}
 =& \frac{\alpha}{2\pi}\int_{0}^{\infty}\frac{{\rm d}s}{s}\,{\rm e}^{-i\frac{m^2}{e}s}\frac{ab}{(a^2+b^2)^2}
 \biggl\{\frac{abs^2}{\sinh^2(as)\sin^2(bs)} \nonumber\\
 &\hspace*{2cm}-\Bigl[1-2\frac{b^2}{a}s\coth(as)+\frac{2}{a^2}\frac{a^4+b^4}{a^2+b^2}\Bigr]\frac{as\cot(bs)}{\sinh^2(as)} + (a\leftrightarrow ib)
 \biggr\},
\nonumber\\
 \frac{\partial^2{\cal L}}{\partial{\cal F}\partial{\cal G}}
 =&\frac{\alpha}{2\pi}\int_{0}^{\infty}\frac{{\rm d}s}{s}\,{\rm e}^{-i\frac{m^2}{e}s}\frac{\cal G}{(a^2+b^2)^2}
 \biggl\{ -\frac{1}{2}\frac{(a^2-b^2)s^2}{\sinh^2(as)\sin^2(bs)} \nonumber\\
 &\hspace*{2cm}+\Bigl[1-2as\coth(as)+\frac{1}{b^2}\frac{(a^2-b^2)^2}{a^2+b^2}\Bigr]\frac{bs\cot(bs)}{\sinh^2(as)} + (a\leftrightarrow ib)
\biggr\}. \label{eq:eq1-5}
\end{align}

Remarkably, for ${\cal G}=0$, i.e., either for orthogonal electric and magnetic fields, or for purely electric or magnetic fields, respectively,
the propertime integrations in \Eqref{eq:eq1-5} can even be performed explicitly
(cf. also \cite{Tsai:1975iz,Dittrich:2000zu,Karbstein:2013ufa}).
Employing integrations by parts, all integrations can be expressed in terms of the elementary integrals
(formulae 3.381.4 and 3.551.3 of \cite{Gradshteyn})
$\int_0^\infty\frac{{\rm d}s}{s}\, s^{\nu}\, {\rm e}^{-\beta s} = \beta^{-\nu}\,\Gamma(\nu)$
and $\int_0^\infty\frac{{\rm d}s}{s}\, s^{\nu}\, {\rm e}^{-\beta s} \coth(s) = \bigl[2^{(1-\nu)}\zeta(\nu,\tfrac{\beta}{2})-\beta^{-\nu}\bigr]\Gamma(\nu)$,
valid for $\Re(\beta)>0$ and under
certain conditions on $\nu$, which are rendered irrelevant upon combination of these integrals in \Eqref{eq:eq1-5}.
In result, the nonvanishing derivatives in this limit are
\begin{align}
 \frac{\partial{\cal L}}{\partial{\cal F}}\bigg|_{{\cal G}=0}
 &=\frac{\alpha}{\pi}\biggl\{4\zeta'(-1,\chi)-\chi\bigl[2\zeta'(0,\chi)-\ln(\chi)+\chi\bigr]-\frac{1}{3}\ln(\chi)-\frac{1}{6}\biggr\},
 \nonumber\\
 \frac{\partial^2{\cal L}}{\partial{\cal F}^2}\bigg|_{{\cal G}=0}
 &=\frac{1}{2{\cal F}}\frac{\alpha}{\pi}\biggl\{\frac{1}{3}-\chi\bigl[2\zeta'(0,\chi)+\ln(\chi)+2\chi\bigl(1-\psi(\chi)\bigr)-1\bigr]\biggr\},
 \nonumber\\
 \frac{\partial^2{\cal L}}{\partial{\cal G}^2}\bigg|_{{\cal G}=0}
 &= \frac{1}{2{\cal F}}\frac{\alpha}{\pi}\biggl\{4\zeta'(-1,\chi)-\chi\bigl[2\zeta'(0,\chi)-\ln(\chi)+\chi\bigr]-\frac{1}{6}\bigl[2\psi(\chi)+\chi^{-1}+1\bigr]\biggr\},
\end{align}
with
\begin{equation}
 \chi = \frac{m^2}{2e\sqrt{2|{\cal F}|}}\times
 \begin{cases}
 1 \quad\text{for}\quad {\cal F}\geq0 \\
 i \quad\text{for}\quad {\cal F}\leq0
 \end{cases},
\end{equation}
while $\frac{\partial{\cal L}}{\partial{\cal G}}\big|_{{\cal G}=0}=\frac{\partial^2{\cal L}}{\partial{\cal F}\partial{\cal G}}\big|_{{\cal G}=0}=0$.
Here, $\psi(\chi)=\frac{\rm d}{{\rm d}\chi}\ln\Gamma(\chi)$ is the Digamma function and $\zeta(s,\chi)$ is the Hurwitz zeta function; $\zeta'(s,\chi)=\partial_s\zeta(s,\chi)$.
Hence, within the above restrictions, for ${\cal G}=0$ explicit analytical insights into the photon polarization tensor in background fields of arbitrary field strengths are possible.

Another important parameter regime is the limit of generic weak background fields.
Counting ${\cal O}(a)\sim {\cal O}(b)
\sim{\cal O}(|\vec{B}|)\sim{\cal O}(|\vec{E}|)\sim{\cal O}(\epsilon)$ and expanding \Eqref{eq:eq1-5} in powers of $\frac{e\epsilon}{m^2}\ll1$,
the propertime integrations can be performed and all prefactors can be determined explicitly.
In particular at leading nonvanishing order in such an expansion we find
\begin{equation}
 \left\{\begin{array}{c} \frac{\partial{\cal L}}{\partial{\cal F}} \\ \frac{\partial{\cal L}}{\partial{\cal G}} \end{array}\right\}  = \frac{\alpha}{\pi}\frac{1}{45}\Bigl(\frac{e}{m^2}\Bigr)^2
 \left\{\begin{array}{c} 4{\cal F} \\7{\cal G} \end{array}\right\}+{\cal O}\bigl((\tfrac{e\epsilon}{m^2})^4\bigr)\,,
\end{equation}
wherefrom -- of course -- also $\frac{\partial^2{\cal L}}{\partial{\cal F}^2}$, $\frac{\partial^2{\cal L}}{\partial{\cal G}^2}$ and $\frac{\partial^2{\cal L}}{\partial{\cal F}\partial{\cal G}}$ can be straightforwardly inferred.
Obviously, the leading contributions to the photon polarization tensor~\eqref{eq:Pi} are of ${\cal O}\bigl((\frac{e\epsilon}{m^2})^2\bigr)$.
As our approach neglects terms $\sim\upsilon^2\,{\cal O}\bigl((\frac{\upsilon}{m})^2\bigr)$,
and the vacuum polarization tensor at zero field scales as $\sim (k^2g^{\rho\sigma}-k^{\rho}k^{\sigma})\,{\cal O}\bigl(\frac{k^2}{m^2}\bigr)\sim\upsilon^2\,{\cal O}\bigl((\frac{\upsilon}{m})^2\bigr)$, no zero field contributions show up.

Finally, we focus on the special case of orthogonal electric and magnetic fields of the same amplitude, i.e., $\vec{E}={\cal E}\vec{e}_E$ and $\vec{B}={\cal E}\vec{e}_B$, with $\vec{e}_E\cdot\vec{e}_B=0$,
which is of major interest when aiming at describing realistic laser backgrounds.
In this limit we have ${\cal F}={\cal G}=0$,
implying that $\frac{\partial{\cal L}}{\partial{\cal F}} = \frac{\partial{\cal L}}{\partial{\cal G}} = \frac{\partial^2{\cal L}}{\partial{\cal F}\partial{\cal G}}=0$.
The only nonvanishing coefficients are
\begin{equation}
 \left\{\begin{array}{c} \frac{\partial^2{\cal L}}{\partial{\cal F}^2} \\ \frac{\partial^2{\cal L}}{\partial{\cal G}^2} \end{array}\right\}  = \frac{\alpha}{\pi}\frac{1}{45}\Bigl(\frac{e}{m^2}\Bigr)^2
 \left\{\begin{array}{c} 4 \\7 \end{array}\right\}.
\end{equation}
This significant simplification is due to the fact that for ${\cal F}={\cal G}=0$ the field dependent higher order contributions to the photon polarization tensor scale as $\sim m^2[(\frac{e\epsilon}{m^2})^{2}{\cal O}(\frac{\upsilon^2}{m^2})]^{n}$, with $n\in\mathbb{N}^+$ (cf. above).
The terms with $n\geq2$ are not accounted for when extracting the photon polarization tensor from the effective Lagrangian~\eqref{eq:effL}, which inherently neglects contributions $\sim\upsilon^2\,{\cal O}\bigl((\frac{\upsilon}{m})^2\bigr)$.
Hence, in this limit the vacuum current vanishes, $j^\sigma(k) = 0$, and the photon polarization tensor is given by
\begin{equation}
 \Pi^{\rho\sigma}(k,k') = \frac{\alpha}{\pi}\frac{1}{45}\Bigl(\frac{e}{m^2}\Bigr)^2\,
  \int_x{\rm e}^{i(k+k')x} \Bigl[ 4\,(k\hat F)^\rho  (k'\hat F)^\sigma + 7\,(k{}^*\hat F)^\rho (k'{}^*\hat F)^\sigma \Bigr]{\cal E}^2 \,, \label{eq:Picrossed}
\end{equation}
where we made use of the definition ${\cal E}\hat F^{\mu\nu}\equiv F^{\mu\nu}$, i.e., scaled out the local electric field amplitude $\cal E$, rendering the
tensor structure in the squared brackets independent of $\cal E$. In turn this tensor structure only depends on the background field alignment
to be characterized by $e^\mu_E\equiv(0,\vec{e}_E)$, $e^\mu_B\equiv(0,\vec{e}_B)$, $\hat\kappa^\mu\equiv(1,\vec{e}_E\times\vec{e}_B)$,
and the probe photons' {\it in-} and {\it outgoing} wave vectors $k'^\mu=(\omega',\vec{k}')$ and $k^\mu=(\omega,\vec{k})$.

A particularly convenient representation of the four vectors in \Eqref{eq:Picrossed} is
\begin{align}
 (k\hat F)^{\mu}&=(k\hat\kappa)e^\mu_E-(k e_E)\hat\kappa^\mu\,, \nonumber\\
 (k{}^*\hat F)^{\mu}&=(k\hat\kappa)e^\mu_B-(k e_B)\hat\kappa^\mu\,.
\end{align}

Assuming the background electromagnetic field to correspond to a plane wave or laser field, the unit vector $\hat{\vec{\kappa}}$ points in the propagation direction of the laser field.
Subsequently, we specialize on a globally fixed $\hat\kappa^\mu$, i.e., the laser's propagation direction does not vary with position and time.
Correspondingly, $\hat\kappa^\mu$ provides a global
reference direction with respect to which the probe photon momenta can be unambiguously decomposed into parallel and perpendicular components,
\begin{equation}
 k^\mu=k_\parallel^\mu+k_\perp^\mu\,,\quad k_\parallel^\mu=(\omega,\vec{k}_\parallel)\,, \quad k_\perp^\mu=(0,\vec{k}_\perp)\,, \label{eq:k}
\end{equation}
with $\vec{k}_\parallel\equiv(\vec{k}\cdot\hat{\vec{\kappa}})\hat{\vec{\kappa}}$ and $\vec{k}_\perp=\vec{k}-\vec{k}_\parallel$.
Let us emphasize that even for fixed $\hat\kappa^\mu$, the orientations of the electric $e^\mu_E$ and magnetic $e^\mu_B$ fields may still depend on space and time, as it is, e.g., the case for circularly polarized electromagnetic fields.

In order to make the subsequent considerations as transparent as possible, without loss of generality we use coordinates where $\hat\kappa^\mu\equiv(1,\vec{e}_{\rm z})$.
The directions of the electric and magnetic fields can then be parameterized by $\vec{e}_E=(\cos\phi,\sin\phi,0)$ and $\vec{e}_B=\vec{e}_E|_{\phi\to\phi+\frac{\pi}{2}}$, with angle parameter $\phi$.
For a linearly polarized laser beam we have $\phi=\text{const}.$, while for circular polarization the parameter $\phi$ generically varies as a function of space and time.
In these coordinates, we have
\begin{equation}
 (k\hat F)^{\mu} = \varepsilon_1^\mu(k)\cos\phi + \varepsilon_2^\mu(k)\sin\phi \quad\text{and}\quad (k{}^*\hat F)^{\mu}=(k\hat F)^{\mu}\big|_{\phi\to\phi+\frac{\pi}{2}}\,,
\end{equation}
with
$\varepsilon_1^\mu(k)\equiv(-k_{\rm x},k_{\rm z}-\omega,0,-k_{\rm x})$ and $\varepsilon_2^\mu(k)\equiv(-k_{\rm y},0,k_{\rm z}-\omega,-k_{\rm y})$.
Note that these four vectors fulfill
$\varepsilon_{1}(k)\varepsilon_2(k)=k\varepsilon_{1}(k)=k\varepsilon_{2}(k)=0$ and $\varepsilon_{1}(k)\varepsilon_1(k)=\varepsilon_{2}(k)\varepsilon_2(k)=(k\hat\kappa)^2$.

\begin{figure}
 \centering
  \includegraphics[width=0.7\textwidth]{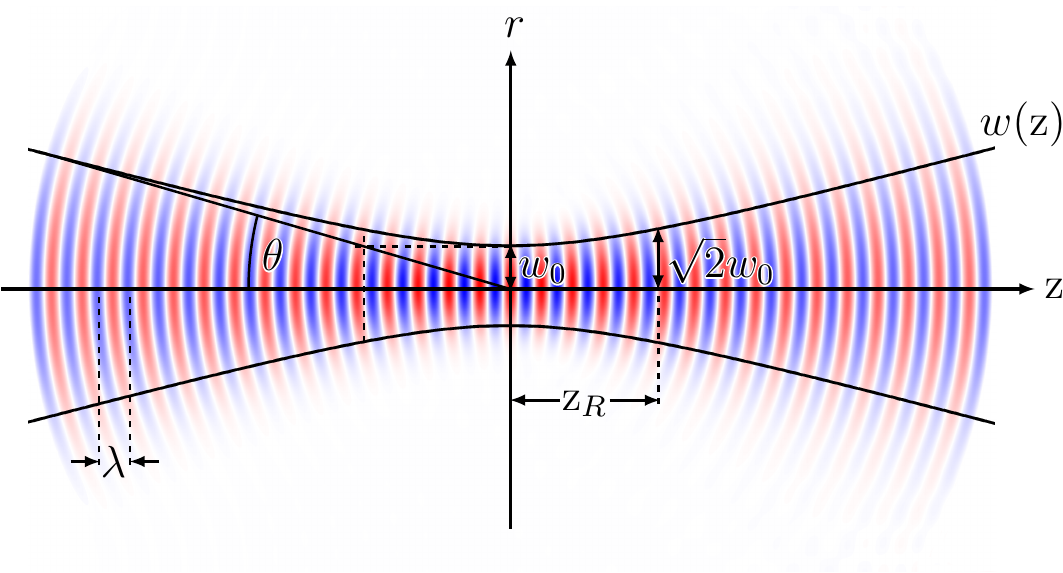} 
\caption{Schematic depiction of a focused Gaussian laser beam (wavelength $\lambda$) propagating along $\rm z$. The beam focus is at ${\rm z}=0$, where the beam diameter is $2w_0$.
The beam is rotationally symmetric about the $\rm z$ axis; $r=\sqrt{{\rm x}^2+{\rm y}^2}$.
The Rayleigh range ${\rm z}_R$ is the distance from the focus (along $\rm z$) for which the beam radius is increased by a factor of $\sqrt{2}$, i.e., $w(\pm{\rm z}_R)=\sqrt{2}w_0$.
The diffraction angle is $\theta\simeq\frac{w_0}{{\rm z}_R}$.}
\label{fig:Gauss}
\end{figure}

Using the conventional notations, the electromagnetic field amplitude profile of a linearly polarized, focused Gaussian laser pulse (cf. Fig.~\ref{fig:Gauss}) in the
paraxial approximation (cf. also below) can be represented as
\begin{equation}
 {\cal E}(r,{\rm z},t)={\cal E}_0{\rm e}^{-\frac{({\rm z}-t)^2}{(\tau/2)^2}}\frac{w_0}{w({\rm z})} {\rm e}^{-(\frac{r}{w({\rm z})})^2}
\cos\bigl(\Phi(r,{\rm z},t)\bigr), \label{eq:env}
\end{equation}
with
\begin{equation}
 \Phi(r,{\rm z},t)=\Omega({\rm z}-t)+\tfrac{\Omega r^2}{2R({\rm z})}-\arctan\bigl(\tfrac{\rm z}{{\rm z}_R}\bigr)+\varphi_0\,, \label{eq:mod}
\end{equation}
i.e., it generically decomposes into an overall pulse envelope and a modulation term $\sim\cos(\Phi)$.
The field or polarization orientation, respectively, is fixed by a particular choice of the angle parameter $\phi=\text{const}.$

The adjustable parameters in Eqs.~\eqref{eq:env} and \eqref{eq:mod} are the peak field strength ${\cal E}_0$, the pulse duration $\tau$, the laser frequency $\Omega=\frac{2\pi}{\lambda}$, the beam's waist size $w_0$ in the laser focus at ${\rm z}=0$ and its Rayleigh range ${\rm z}_R=\frac{\pi w_0^2}{\lambda}$.
Moreover, $w({\rm z})=w_0\sqrt{1+(\frac{\rm z}{{\rm z}_R})^2}$ describes the widening of the beam's transverse extent as a function of $\rm z$. The Rayleigh range is the distance from the focus (along $\rm z$) for which the beam radius is increased by a factor of $\sqrt{2}$, i.e., $w(\pm{\rm z}_R)=\sqrt{2}w_0$, and correspondingly the beam cross section is increased by a factor of two as compared to the beam waist.
Finally, $R({\rm z})={\rm z}\bigl[1+(\frac{{\rm z}_R}{\rm z})^2\bigr]$ is the radius of curvature of the wavefronts, and the term $\arctan\bigl(\tfrac{\rm z}{{\rm z}_R}\bigr)$ in \Eqref{eq:mod} accounts for the Gouy phase shift; cf., e.g., \cite{Siegman}.
In addition we accounted for a phase $\varphi_0$.
The dominant frequency scale governing this type of field configuration is $\Omega$, which in our approach, counting ${\cal O}(\Omega)={\cal O}(\upsilon)$, is constrained to small values of $\frac{\Omega}{m}\ll1$ (cf. above).

An alternative representation of Eqs.~\eqref{eq:env} and \eqref{eq:mod}, which is particularly convenient for our purposes, is obtained by elementary manipulations and reads
\begin{equation}
 {\cal E}(r,{\rm z},t)={\cal E}_0\,{\rm e}^{-\frac{({\rm z}-t)^2}{(\tau/2)^2}} \Bigl(\frac{w_0}{w({\rm z})}\Bigr)^2
 \,\frac{1}{2}\sum_{l=\pm1}\Bigl(1-il\frac{{\rm z}}{{\rm z}_R}\Bigr) {\rm e}^{-(1-il\frac{\rm z}{{\rm z}_R})(\frac{r}{w({\rm z})})^2}{\rm e}^{il\Omega({\rm z}-t)+il\varphi_0}. \label{eq:Elin}
\end{equation}
The (leading order) paraxial approximation adopted here is valid for small diffraction angles $\theta\simeq\frac{w_0}{{\rm z}_R}$, and neglects terms of ${\cal O}(\theta)$ and ${\cal O}(\frac{1}{\tau\Omega})$.
It can be systematically improved to account for higher order contributions in the parameter $\theta$ \cite{Davis:1979,Barton:1989,Salamin:2002dd}.
In particular, note that at leading order in the paraxial approximation, the electric and magnetic field vectors are still orthogonal to each other and to the laser propagation direction; cf. also our assumptions below \Eqref{eq:k}.

To describe a circularly polarized Gaussian laser pulse we use the following electric and magnetic fields,
\begin{align}
 \!\!\vec{E}_\pm(r,{\rm z},t)&={\cal E}_0\,{\rm e}^{-\frac{({\rm z}-t)^2}{(\tau/2)^2}}\frac{w_0}{w({\rm z})} {\rm e}^{-(\frac{r}{w({\rm z})})^2}
 \bigl[\cos\bigl(\Phi(r,{\rm z},t)\bigr)\vec{e}_{\rm x}\pm\sin\bigl(\Phi(r,{\rm z},t)\bigr)\vec{e}_{\rm y}\bigr]  \nonumber\\
 &={\cal E}_0\,{\rm e}^{-\frac{({\rm z}-t)^2}{(\tau/2)^2}} \Bigl(\frac{w_0}{w({\rm z})}\Bigr)^2
 \,\frac{1}{2}\sum_{l=\pm1}\Bigl(1-il\frac{{\rm z}}{{\rm z}_R}\Bigr) {\rm e}^{-(1-il\frac{\rm z}{{\rm z}_R})(\frac{r}{w({\rm z})})^2}{\rm e}^{il\Omega({\rm z}-t)+il\varphi_0} \bigl(\vec{e}_{\rm x}\mp i^l\vec{e}_{\rm y}\bigr) \label{eq:Ecirc}
\end{align}
and $\vec{B}_\pm=\vec{E}_\pm|_{\Phi\to\Phi+\frac{\pi}{2}}$,
where $+/-$ stand for right-/lefthanded circular polarizations.

Equation~\eqref{eq:Elin} and the second line of \eqref{eq:Ecirc} are particularly suited for an explicit evaluation of the photon polarization tensor in high-intensity laser backgrounds.
Noteworthily, for these field configurations all integrations in \Eqref{eq:Picrossed} can be performed analytically.
Our result for the photon polarization tensor in a linearly polarized Gaussian laser pulse reads
\begin{multline}
 \Pi^{\rho\sigma}(k,k') = \frac{\alpha}{45}\frac{\pi}{4}\Bigl(\frac{e{\cal E}_0}{m^2}\Bigr)^2 w_0^2\,{\rm z}_R\,\frac{\tau}{2}\,
  \Bigl[ 4\,(k\hat F)^\rho  (k'\hat F)^\sigma + 7\,(k{}^*\hat F)^\rho (k'{}^*\hat F)^\sigma \Bigr] \\
 \times
 {\rm e}^{-\frac{1}{8}(\frac{\tau}{2})^2(\omega+\omega')^2}
 \biggl[2\,{\rm e}^{-2\frac{{\rm z}_R^2[(k+k')\hat\kappa]^2}{w_0^2(k_\perp+k_\perp')^2}}\frac{{\rm e}^{-\frac{1}{8}w_0^2(k_\perp+k_\perp')^2}}{\sqrt{w_0^2(k_\perp+k_\perp')^2}}  \\
 + \sqrt{\frac{\pi}{2}}\sum_{l=\pm1}{\rm e}^{-\frac{1}{2}(\frac{\tau}{2})^2 \Omega[\Omega+l(\omega+\omega')]+il2\varphi_0}\,{\rm e}^{-l{\rm z}_R(k+k')\hat\kappa} 
\ \Theta\Bigl(l{\rm z}_R(k+k')\hat\kappa-\tfrac{1}{8}w_0^2(k_\perp+k_\perp')^2\Bigr) \biggr], \label{eq:PIlin}
\end{multline}
where $\Theta(.)$ is the Heaviside function, and we used the shorthand notations $(k+k')\hat\kappa=(k+k')_\mu\hat\kappa^\mu$ and $(k_\perp+k_\perp')^2=(\vec{k}_\perp+\vec{k}_\perp')^2$.
In this case the electric and magnetic fields point in fixed directions and the tensor structures in \Eqref{eq:Picrossed} are independent of space and time and can be factored out.

The analogous expression for circularly polarized Gaussian laser pulses is
\begin{multline}
  \Pi^{\rho\sigma}_\pm(k,k')
 = \frac{\alpha}{45}\frac{\pi}{4}\Bigl(\frac{e{\cal E}_0}{m^2}\Bigr)^2w_0^2\,{\rm z}_R\,\frac{\tau}{2}\, {\rm e}^{-\frac{1}{8}(\frac{\tau}{2})^2(\omega+\omega')^2}  \\
\times\!\biggl\{22\,{\rm e}^{-2\frac{{\rm z}_R^2[(k+k')\hat\kappa]^2}{w_0^2(k_\perp+k_\perp')^2}} \frac{{\rm e}^{-\frac{1}{8}w_0^2(k_\perp+k_\perp')^2}}{\sqrt{w_0^2(k_\perp+k_\perp')^2}}
 \bigl[\varepsilon_1^\rho(k)\varepsilon_1^\sigma(k')+ \varepsilon_2^\rho(k)\varepsilon_2^\sigma(k')\bigr]\\
-3\sqrt{\frac{\pi}{2}}\sum_{l=\pm1}{\rm e}^{-\frac{1}{2}(\frac{\tau}{2})^2 \Omega[\Omega+l(\omega+\omega')]+il2\varphi_0}\,{\rm e}^{-l{\rm z}_R(k+k')\hat\kappa}
 \ \Theta\Bigl(l{\rm z}_R(k+k')\hat\kappa-\tfrac{1}{8}w_0^2(k_\perp+k_\perp')^2\Bigr) \\
 \times\!\Bigl[\bigl[\varepsilon_1^\rho(k)\varepsilon_1^\sigma(k')-\varepsilon_2^\rho(k)\varepsilon_2^\sigma(k')\bigr]
 \pm(-i)^l\bigl[\varepsilon_1^\rho(k)\varepsilon_2^\sigma(k')  + \varepsilon_2^\rho(k)\varepsilon_1^\sigma(k')\bigr]\Bigr]\!\biggr\},\!\! \label{eq:PIcirc}
\end{multline}
where the subscript $\pm$ refers to the results for right-/lefthanded circular polarizations (cf. above).
As for circular polarization the electric and magnetic field vectors rotate as a function of position and time, i.e., $\phi\to\pm\Phi(r,{\rm z},t)$, in this case
we find it more convenient to represent the tensor structures in terms of the four vectors $\varepsilon_1^\mu$ and $\varepsilon_2^\mu$.

Equations~\eqref{eq:PIlin} and \eqref{eq:PIcirc} are made up of three generic contributions: One term that is independent of the laser frequency $\Omega$, and two $\Omega$ dependent terms.
The structure of the $\Omega$ independent term differs significantly from the others.
In the limit of an infinitely long pulse duration $\tau$, the first term corresponds to an elastic photon scattering process with $\omega=-\omega'$,
and the latter terms can be identified with inelastic processes to be associated with the absorption (emission) of two laser photons of frequency $\Omega$, i.e., $\omega=-(\omega'\pm2\Omega)$;
note that $\lim_{\tau\to\infty}\frac{\tau}{2}\,{\rm e}^{-(\frac{\tau}{2})^2\phi^2} = \sqrt{\pi}\,\delta(\phi)$.

This generic structure is well known from the photon polarization tensor in plane wave backgrounds \cite{Baier:1975ff,Becker:1974en}.
In particular at order $(\frac{e\epsilon}{m^2})^2$ and for soft electromagnetic fields, the plane wave photon polarization tensor comprises the processes of zero and $\pm2\Omega$ photon absorption/emission from the background field \cite{Gies:2014jia}.

\section{Conclusions} \label{sec:conclusions}

In this paper we have derived expressions for the photon polarization tensor in inhomogeneous electromagnetic background fields,
valid for electromagnetic fields which vary on scales much larger than the Compton wavelength and Compton time of the electron, respectively.
While for homogeneous backgrounds the photon polarization tensor in momentum space can be fully expressed in terms of the momentum transfer $(k+k')^\mu$,  
for inhomogeneous backgrounds, it generically mediates between two distinct in- and outgoing four momenta $k'^\mu$ and $k^\mu$.
In contrast to previous calculations, we have accounted for the full tensor structure spanned by $k'^\mu$ and $k^\mu$, rendering our result fully consistent with the Ward identity.

For most generic electromagnetic field configurations and arbitrary field strengths, the resulting expression can be written in terms of a Fourier integral over a single propertime integral.
In the special case of either orthogonal electric and magnetic fields, or purely electric or magnetic fields, the propertime integration can even be performed explicitly.
Besides, the propertime integrations can also be performed explicitly in the limit of generic weak electromagnetic fields.

A central outcome of our analysis is compact explicit expressions for the polarization tensor in realistic laser backgrounds, namely for focused linearly and circularly polarized Gaussian laser pulses which are well described by the paraxial approximation, i.e., for small diffraction angles, $\theta\ll1$, and large values for the product of the laser pulse duration and frequency $\tau\Omega\gg1$.
Other restrictions for our result to be trustworthy are $\tau_c\Omega\ll1$ and $\frac{I}{I_{\rm cr}}(\tau_c\Omega)^2\ll1$, with laser peak intensity $I={\cal E}_0^2$; recall that the electron Compton time (length) is $\tau_c=1.29\cdot10^{-21}{\rm s}$ ($\lambda_c=3.86\cdot10^{-13}{\rm m}$ ), and the critical intensity is $I_{\rm cr}\equiv(\frac{m^2}{e})^2=4.68\cdot10^{29}\frac{\rm W}{{\rm cm}^2}$.
All these conditions are typically met for contemporary optical high-intensity laser systems delivering multicycle laser pulses.
Moreover, the probe photon frequencies have to fulfill $\{\tau_c\,\omega,\tau_c\,\omega'\}\ll1$, and their momenta $\{\lambda_c|\vec{k}|,\lambda_c|\vec{k}'|\}\ll1$,
which is certainly true for the whole range from optical to soft X-ray frequency photons available in the laboratory.

In particular, the latter results are of utmost importance for the investigation of quantum vacuum nonlinearities in realistic high-intensity laser experiments, facilitating, e.g., a deep and thorough analysis of vacuum birefringence \cite{Heinzl:2006xc} or quantum reflection \cite{Gies:2013yxa}.

\section*{Acknowledgments}

It is a great pleasure to thank H.~Gies for many stimulating discussions and support, as well as for valuable comments on the manuscript.
We are particularly indebted to Maria~Reuter for creating Fig.~\ref{fig:Gauss}, and would like to thank A. Shabad for valuable comments on the manuscript.
FK is grateful to M.~Zepf for various helpful discussions, and would like to thank M.~C.~Kaluza and J.~Polz for access to their private libraries.
FK acknowledges support by the DFG (SFB-TR18), and RS acknowledges support by the Ministry of Education and Science of the Republic of Kazakhstan.


\begin{thebibliography}{10}\setlength{\itemsep}{-0.5mm}

\bibitem{Euler:1935zz} 
  H.~Euler and B.~Kockel,
  Naturwiss.\  {\bf 23}, 246 (1935).

\bibitem{Heisenberg:1935qt} 
  W.~Heisenberg and H.~Euler,
  Z.\ Phys.\  {\bf 98}, 714 (1936), 
  an English translation is available at [physics/0605038].

\bibitem{Weisskopf}
V.~Weisskopf, 
Kong.\ Dans.\ Vid.\ Selsk., Mat.-fys.\ Medd.\ {\bf XIV}, 6 (1936).

\bibitem{Akhmadaliev:1998zz} 
  S.~Z.~Akhmadaliev, {\it et al.}, 
  Phys.\ Rev.\ C {\bf 58}, 2844 (1998);
  S.~Z.~Akhmadaliev, {\it et al.}, 
  Phys.\ Rev.\ Lett.\  {\bf 89}, 061802 (2002)
  [hep-ex/0111084].

\bibitem{Dittrich:2000zu} 
  W.~Dittrich and H.~Gies,
  Springer Tracts Mod.\ Phys.\  {\bf 166}, 1 (2000).

\bibitem{Marklund:2008gj} 
  M.~Marklund and J.~Lundin,
  Eur.\ Phys.\ J.\ D {\bf 55}, 319 (2009)
  [arXiv:0812.3087 [hep-th]].

\bibitem{Dunne:2008kc} 
  G.~V.~Dunne,
  Eur.\ Phys.\ J.\ D {\bf 55}, 327 (2009)
  [arXiv:0812.3163 [hep-th]].
  
\bibitem{Heinzl:2008an} 
  T.~Heinzl and A.~Ilderton,
  Eur.\ Phys.\ J.\ D {\bf 55}, 359 (2009)
  [arXiv:0811.1960 [hep-ph]]; .
  
\bibitem{DiPiazza:2011tq} 
  A.~Di Piazza, C.~Muller, K.~Z.~Hatsagortsyan and C.~H.~Keitel,
  Rev.\ Mod.\ Phys.\  {\bf 84}, 1177 (2012)
  [arXiv:1111.3886 [hep-ph]].

\bibitem{Toll:1952}
J.~S.~Toll,
Ph.D. thesis, Princeton Univ., 1952 (unpublished).

\bibitem{Baier}
R.~Baier and P.~Breitenlohner, 
{Act.~Phys.~Austriaca} {\bf 25}, 212 (1967); 
{Nuov.~Cim.~B}\ {\bf 47} 117 (1967).
  
\bibitem{BialynickaBirula:1970vy} 
  Z.~Bialynicka-Birula and I.~Bialynicki-Birula,
  Phys.\ Rev.\ D {\bf 2}, 2341 (1970).

\bibitem{Adler:1971wn}
  S.~L.~Adler,
  {Annals Phys.}\  {\bf 67}, 599 (1971).

\bibitem{Heinzl:2006xc} 
  T.~Heinzl, B.~Liesfeld, K.~-U.~Amthor, H.~Schwoerer, R.~Sauerbrey and A.~Wipf,
  Opt.\ Commun.\  {\bf 267}, 318 (2006)
  [hep-ph/0601076].

\bibitem{Dinu:2013gaa} 
  V.~Dinu, T.~Heinzl, A.~Ilderton, M.~Marklund and G.~Torgrimsson,
  Phys.\ Rev.\ D {\bf 89}, 125003 (2014)
  [arXiv:1312.6419 [hep-ph]]; 
  Phys.\ Rev.\ D {\bf 90}, 045025 (2014)
  [arXiv:1405.7291 [hep-ph]].
  
\bibitem{Cantatore:2008zz} 
  G.~Cantatore [PVLAS Collaboration],
  Lect.\ Notes Phys.\  {\bf 741}, 157 (2008); 
  E.~Zavattini {\it et al.}  [PVLAS Collaboration],
  Phys.\ Rev.\ D {\bf 77}, 032006 (2008)
  [arXiv:0706.3419 [hep-ex]]; 
  F.~Della Valle, U.~Gastaldi, G.~Messineo, E.~Milotti, R.~Pengo, L.~Piemontese, G.~Ruoso and G.~Zavattini,
  arXiv:1301.4918 [quant-ph].

\bibitem{Berceau:2011zz} 
  P.~Berceau, R.~Battesti, M.~Fouche and C.~Rizzo,
  Can.\ J.\ Phys.\  {\bf 89}, 153 (2011);
  P.~Berceau, M.~Fouche, R.~Battesti and C.~Rizzo,
  Phys.\ Rev.\ A, {\bf 85}, 013837 (2012)
  [arXiv:1109.4792 [physics.optics]];
  A.~Cadene, P.~Berceau, M.~Fouche, R.~Battesti and C.~Rizzo,
  Eur.\ Phys.\ J.\ D {\bf 68}, 16 (2014)
  [arXiv:1302.5389 [physics.optics]].

\bibitem{Zavattini:2008cr} 
  G.~Zavattini and E.~Calloni,
  Eur.\ Phys.\ J.\ C {\bf 62}, 459 (2009)
  [arXiv:0812.0345 [physics.ins-det]].

\bibitem{Dobrich:2009kd} 
  B.~Dobrich and H.~Gies,
  Europhys.\ Lett.\  {\bf 87}, 21002 (2009)
  [arXiv:0904.0216 [hep-ph]].

\bibitem{Grote:2014hja} 
  H.~Grote,
  arXiv:1410.5642 [physics.ins-det].

\bibitem{Karplus:1950zz} 
  R.~Karplus and M.~Neuman,
  Phys.\ Rev.\  {\bf 83}, 776 (1951).
  
\bibitem{Sauter:1931zz} 
  F.~Sauter,
  Z.\ Phys.\  {\bf 69}, 742 (1931).
  
\bibitem{Schwinger:1951nm} 
  J.~S.~Schwinger,
  Phys.\ Rev.\  {\bf 82}, 664 (1951).

\bibitem{King:2013am} 
  B.~King, A.~Di Piazza and C.~H.~Keitel,
 Nature Photon.\ {\bf 4}, 92 (2010) 
  [arXiv:1301.7038 [physics.optics]]; 
Phys.\ Rev.\ A {\bf 82}, 032114 (2010)
  [arXiv:1301.7008 [physics.optics]].

\bibitem{Tommasini:2010fb} 
  D.~Tommasini and H.~Michinel,
  Phys.\ Rev.\ A {\bf 82}, 011803 (2010)
  [arXiv:1003.5932 [hep-ph]].
  
\bibitem{Hatsagortsyan:2011}
K.~Z.~Hatsagortsyan and G.~Y.~Kryuchkyan,
 Phys.\ Rev.\ Lett. {\bf 107}, 053604 (2011).

\bibitem{King:2012aw} 
  B.~King and C.~H.~Keitel,
  New J.\ Phys.\  {\bf 14}, 103002 (2012)
  [arXiv:1202.3339 [hep-ph]].

\bibitem{Gies:2013yxa} 
  H.~Gies, F.~Karbstein and N.~Seegert,
  New J.\ Phys.\  {\bf 15}, 083002 (2013)
  [arXiv:1305.2320 [hep-ph]];
  H.~Gies, F.~Karbstein and N.~Seegert,
  arXiv:1412.0951 [hep-ph].

\bibitem{Gies:2014jia} 
  H.~Gies, F.~Karbstein and R.~Shaisultanov,
  Phys.\ Rev.\ D {\bf 90}, no. 3, 033007 (2014)
  [arXiv:1406.2972 [hep-ph]].

\bibitem{DiPiazza:2005jc} 
  A.~Di Piazza, K.~Z.~Hatsagortsyan and C.~H.~Keitel,
  Phys.\ Rev.\ D {\bf 72}, 085005 (2005).

\bibitem{Fedotov:2006ii} 
  A.~M.~Fedotov and N.~B.~Narozhny,
  Phys.\ Lett.\ A {\bf 362}, 1 (2007)
  [hep-ph/0604258].
  
\bibitem{BatShab}
  I.~A.~Batalin and A.~E.~Shabad,
  Zh.\ Eksp.\ Teor.\ Fiz.\  {\bf 60}, 894 (1971)
  [Sov.\ Phys.\ JETP\ {\bf 33}, 483 (1971)].

\bibitem{narozhnyi:1968}
  N.~B.~Narozhnyi,
 Zh.\ Eksp.\ Teor.\ Fiz.\ {\bf 55}, 714 (1968)
 [Sov.\ Phys.\ JETP \textbf{28}, 371 (1969)].
  
\bibitem{ritus:1972}
 V.~I.~Ritus,
 Ann.\ Phys.\ {\bf 69}, 555 (1972).

\bibitem{Tsai:1974fa}
  W.~y.~Tsai and T.~Erber,
  {\it Phys.\ Rev.\  D}\ {\bf 10}, 492 (1974).
 
\bibitem{Tsai:1974ap}
  W.~y.~Tsai,
  Phys.\ Rev.\  D {\bf 10}, 2699 (1974).
 
\bibitem{Baier:1974hn} 
  V.~N.~Baier, V.~M.~Katkov and V.~M.~Strakhovenko,
  Zh.\ Eksp.\ Teor.\ Fiz.\  {\bf 68}, 405 (1975)
  [Sov.\ Phys.\ JETP\ {\bf 41}, 198 (1975)].
 
\bibitem{Urrutia:1977xb}
  L.~F.~Urrutia,
  {\it Phys.\ Rev.\  D}\ {\bf 17}, 1977 (1978).
  
\bibitem{Dittrich:2000wz} 
  W.~Dittrich and R.~Shaisultanov,
  Phys.\ Rev.\ D {\bf 62}, 045024 (2000)
  [hep-th/0001171].
  
\bibitem{Schubert:2000yt} 
  C.~Schubert,
  Nucl.\ Phys.\ B {\bf 585}, 407 (2000)
  [hep-ph/0001288].

\bibitem{Karbstein:2013ufa} 
  F.~Karbstein,
  Phys.\ Rev.\ D {\bf 88}, no. 8, 085033 (2013)
  [arXiv:1308.6184 [hep-th]].

\bibitem{Baier:1975ff} 
  V.~N.~Baier, A.~I.~Milshtein and V.~M.~Strakhovenko,
  Zh.\ Eksp.\ Teor.\ Fiz.\  {\bf 69}, 1893 (1975)
  [Sov.\ Phys.\ JETP\ {\bf 42}, 961 (1976)].

\bibitem{Becker:1974en} 
  W.~Becker and H.~Mitter,
   J.\ Phys.\ A:\ Math.\ Gen.\ {\bf 8} 1638 (1975).

\bibitem{Meuren:2013oya} 
  S.~Meuren, C.~H.~Keitel and A.~Di Piazza,
  Phys.\ Rev.\ D {\bf 88}, no. 1, 013007 (2013)
  [arXiv:1304.7672 [hep-ph]].

\bibitem{Gies:2011he} 
  H.~Gies and L.~Roessler,
  Phys.\ Rev.\ D {\bf 84}, 065035 (2011)
  [arXiv:1107.0286 [hep-ph]].
  
\bibitem{Jentschura:2001qr} 
  U.~D.~Jentschura, H.~Gies, S.~R.~Valluri, D.~R.~Lamm and E.~J.~Weniger,
  Can.\ J.\ Phys.\  {\bf 80}, 267 (2002)
  [hep-th/0107135].

\bibitem{Shaisultanov:1997bc} 
  R.~Shaisultanov,
  Phys.\ Rev.\ Lett.\  {\bf 80}, 1586 (1998)
  [hep-ph/9709420].
  
\bibitem{Karbstein:2007be} 
  F.~Karbstein and M.~Thies,
  Phys.\ Rev.\ D {\bf 77}, 025008 (2008)
  [arXiv:0708.3176 [hep-th]];
  F.~Karbstein,
  Phys.\ Rev.\ C {\bf 81}, 045206 (2010)
  [arXiv:0908.2759 [hep-th]].
  
\bibitem{Karbstein:2014fva} 
  F.~Karbstein and R.~Shaisultanov,
  arXiv:1412.6050 [hep-ph].

\bibitem{Shabad:2011hf} 
  A.~E.~Shabad and V.~V.~Usov,
  Phys.\ Rev.\ D {\bf 83}, 105006 (2011)
  [arXiv:1101.2343 [hep-th]].
  
\bibitem{Tsai:1975iz} 
  W.~y.~Tsai and T.~Erber,
  Phys.\ Rev.\ D {\bf 12}, 1132 (1975).
  
\bibitem{Gradshteyn}
I.~S.~Gradshteyn and I.~M.~Ryzhik, \textit{Table of Integrals, Series, and Products}, Fifth Edition, Academic Press, UK (1994).
  
\bibitem{Siegman}
A.~E.~Siegman, \textit{Lasers}, First Edition, University Science Books, USA (1986);
B.~E.~A.~Saleh and M.~C.~Teich, \textit{Fundamentals of Photonics}, First Edition, John Wiley \& Sons, USA (1991).
  
\bibitem{Davis:1979} 
  L.~W.~Davis,
  Phys.\ Rev.\ A {\bf 19}, 3 (1979).
  
\bibitem{Barton:1989}
 J.~P.~Barton and D.~R.~Alexander,
 J. Appl. Phy. {\bf 66}, 2800 (1989).
  
\bibitem{Salamin:2002dd} 
  Y.~I.~Salamin, G.~R.~Mocken and C.~H.~Keitel,
  Phys.\ Rev.\ ST Accel.\ Beams {\bf 5}, 101301 (2002);
  Y.~I.~Salamin, S.~X.~Hu, K.~Z.~Hatsagortsyan and C.~H.~Keitel,
  Phys.\ Rept.\  {\bf 427}, 41 (2006).
  
\bibitem{Gies:2014jia} 
  H.~Gies, F.~Karbstein and R.~Shaisultanov,
  Phys.\ Rev.\ D {\bf 90}, 033007 (2014)
  [arXiv:1406.2972 [hep-ph]].
  
\end{thebibliography}
\end{document}